# LUCF-Net: Lightweight U-shaped Cascade Fusion Network for Medical Image Segmentation

Songkai Sun, Qingshan She, Yuliang Ma, Rihui Li, Yingchun Zhang, *Senior Member, IEEE*

*Abstract*—In this study, the performance of existing U-shaped neural network architectures was enhanced for medical image segmentation by adding Transformer. Although Transformer architectures are powerful at extracting global information, its ability to capture local information is limited due to its high complexity. To address this challenge, we proposed a new lightweight U-shaped cascade fusion network (LUCF-Net) for medical image segmentation. It utilized an asymmetrical structural design and incorporated both local and global modules to enhance its capacity for local and global modeling. Additionally, a multi-layer cascade fusion decoding network was designed to further bolster the network's information fusion capabilities. Validation results achieved on multi-organ datasets in CT format, cardiac segmentation datasets in MRI format, and dermatology datasets in image format demonstrated that the proposed model outperformed other state-of-the-art methods in handling local-global information, achieving an improvement of 1.54% in Dice coefficient and 2.6 mm in Hausdorff distance on multi-organ segmentation. Furthermore, as a network that combines Convolutional Neural Network and Transformer architectures, it achieves competitive segmentation performance with only 6.93 million parameters and 6.6 gigabytes of floating point operations, without the need of pre-training. In summary, the proposed method demonstrated enhanced performance while retaining a simpler model design compared to other Transformer-based segmentation networks.

## I. Introduction

IN medical image analysis, artificial intelligence is considered the de-facto method to build computer-aided diagnosis applications for image segmentation [1]. Among these AI-based applications, image segmentation plays a pivotal role in facilitating disease diagnosis and the formulation of treatment strategies [2]. With the rapid advancements in deep learning methods, Convolutional Neural Network (CNN) and Transformer have emerged as the two main research directions in medical auxiliary analysis in recent years [3]. Both of these methods have distinct advantages, providing new opportunities for tackling intricate medical image segmentation challenges. The U-Net [4], which consists solely of CNN with up-sampling, down-sampling, and skip connections, has demonstrated impressive performance with minimal complexity across different domains. It has particularly excelled in tasks like segmenting multiple organs and skin lesions, solidifying its distinct role in medical image segmentation.

However, challenges persist in the application of CNN for medical image analysis [5]. Medical images often contain a wealth of contextual information that spans large areas, capturing the overall structure, shape, and distribution of the image. This comprehensive view is crucial for precise diagnosis and treatment planning, taking into account factors like the overall layout, size, and spatial relationships within the organ or tissue. By utilizing these distant dependencies, a more accurate and detailed analysis of medical images can be achieved. CNN may have limitations in capturing distant correlations effectively, which could result in overlooking global information and impacting segmentation accuracy. To address this issue, Transformer, a model utilizing self-attention mechanisms, has gained significant recognition. Its excellent long-range dependency modeling capability has been introduced into computer vision, demonstrating remarkable achievements in image segmentation tasks. Compared with CNN, Transformer has certain advantages in some aspects of the medical image field. Firstly, Transformer can capture global dependencies between pixels in the image [6], allowing for better understanding of the overall structure. Secondly, Transformer can offer higher flexibility [7]. Traditional CNN models usually require manual design of the network structure, while Transformer models can adapt to different tasks through simple modifications, such as adding or reducing layers or heads. Therefore, Transformer models are more flexible in handling various visual tasks. Despite these advantages over CNN,

This work was partly supported by the National Natural Science Foundation of China (62371172 and 82301743), the Science and Technology Development Fund of the Macao SAR (0010/2023/ITP1), the University of Macau (SRG2023-00015-ICI), and Zhejiang Provincial Natural Science Foundation of China (No. LZ22F010003). (Corresponding author: Qingshan She and Rihui Li.)

Songkai Sun, Qingshan She and Yuliang Ma are with the School of Automation, Hangzhou Dianzi University, Hangzhou, Zhejiang 310018, China, and also with International Joint Research Laboratory for Autonomous Robotic Systems, Hangzhou, Zhejiang, China (e-mail: qsshe@hdu.edu.cn).

Rihui Li is with the Center for Cognitive and Brain Sciences, and the Department of Electrical and Computer Engineering, University of Macau, Macau SAR, China (e-mail: rihuili@um.edu.mo).

Yingchun Zhang is with the Department of Biomedical Engineering, University of Miami, Coral Gables, Florida, USA (e-mail: y.zhang@miami.edu).

Transformer have one fatal flaw [8]: the computational efficiency of Transformer-based networks is often considerably less than CNN networks, leading to demanding computational costs. Thus, how to efficiently utilize Transformer models becomes a critical issue.

To further enhance the performance of medical image segmentation, researchers have begun to explore methods that combine CNN and Transformer [9]. Through the integration of the strengths of each, there is potential to enhance the handling of complex attributes and long-range dependencies in medical images, ultimately leading to more accurate and reliable segmentation results with reduced model complexity. Nonetheless, earlier research that integrated CNN and Transformer simply merged them together without fundamentally addressing the complexity issue of Transformer networks [10]. In this study, inspired by EdgeViTs [11], an asymmetric CNN-Transformer network is proposed based on local-global feature cascading. It incorporates patch-wise self-attention after down-sampling to accomplish local and global feature extraction while significantly reduces network complexity. By constructing an efficient local-global feature extraction module in the U-shaped network encoder, the local features extracted by CNN are effectively integrated with the global features extracted by Transformer.

This research primarily offers the following key contributions:
1) By incorporating an efficient local-global feature extraction module into the U-shaped network encoder, the local features derived from the CNN seamlessly integrate with the global features extracted by the Transformer.
2) An asymmetric U-shaped network architecture is designed to reduce model complexity. Multi-layer feature fusion is performed in the decoder, and the loss is calculated layer by layer throughout the training process which accelerates the rate of convergence of the network and enhances the network's ability to fuse local and global information.
3) A new combination of multiple loss functions is employed to address the issue of dataset sample imbalance, and the segmentation accuracy is further improved through an online hard sample learning strategy.

The rest of this paper is structured as outlined below. Section II reviews the application of CNN and Transformer models in medical image segmentation. Section III introduces the structure of the proposed lightweight U-shaped cascade fusion network (LUCF-Net) in detail. Section IV introduces the overall experiment, including verification of model performance and complexity and ablation experiments. Section V discusses LUCF-Net. Section VI is the conclusion of this work.

## II. RELATED WORK

### A. CNN-based Networks

Early medical image segmentation methods mostly adopt pure CNN structures. U-Net is undoubtedly a pioneering work in this field. It combines decoders, encoders and skip connections to lay the foundation for U-shaped network architecture. After U-Net was proposed, various methods based on U-Net have been introduced [12]. Diakogiannis et al. [13] used the U-Net encoder/decoder backbone, and combined residual connections [14], hole models, pyramid scene parsing pools and multi-task reasoning to implement the ResUNet-a model, thereby retaining the image segmentation architecture of U-Net while enhancing feature propagation and learning capabilities. By implementing an attention gating module, Thomas et al. [15] were able to enhance U-Net by utilizing feature maps to capture global information and improve long-range dependency modeling. Do et al. [16] combined global and patch-based methods to achieve global information modeling by utilizing multi-level distance features. Guan et al. [17] integrated the concept of dense networks, connecting the features of each decoder layer with preceding encoder layers to achieve more robust feature propagation. In the improved versions such as UNet++ [18], IR-UNet++ [19] and UNet3+ [20], which utilized skip connections, multi-level feature fusion, and up-sampling structures, the information propagation and feature extraction capabilities of the model were further enhanced [19, 20]. In the segmentation of three-dimensional medical images [23], 3D-UNet [24] and VNet [25], based on 3D convolutions, were introduced, enabling medical image segmentation networks to be suitable for volumetric data. The above-mentioned CNN-based methods predominantly employed multi-layer feature fusion, attention mechanisms, and other techniques to compensate for the CNN network's inherent limitation in global modeling ability. As a result, these approaches contribute to performance enhancements to a certain degree.

### B. Transformer-based Networks

The Transformer was initially introduced within the domain of Natural Language Processing (NLP) and is renowned for its exceptional capacity to capture extensive interdependencies. Dsosovitskiy et al. [26] extended the utilization of the Transformer to the realm of computer vision by segmenting images into tokens for use within the Transformer network. This breakthrough greatly enhanced the network's capacity to extract global features. As a pioneering effort, TransUNet [27] integrated the Transformer into the U-shaped network architecture. It not only encoded strong global context by encoding image features as sequences, but also made good use of low-level CNN features through U-Net hybrid network design. Cao et al. [28] further combined the Swin Transformer [29], substituting the decoder and encoder with Transformer networks, generating a pure Transformer U-shaped network to repair the shortcomings of CNN networks in global features. Similarly, DS TransUNet [30] employed dense networks to construct a dense-connected pure Transformer U-shaped network, building upon the foundation of TransUNet. Faced with computational limitations inherent in Transformer, an increasing number of researchers began investigating more efficient Transformer-based U-Net architectures. Huang et al. [31] proposed MISSFormer, which redesigned the feed-forward network within the encoder structure, facilitating more efficient local and global contextual feature extraction. Reza et al. [32] introduced DAEFormer, which redefined the self-attention mechanism and skip connection paths. This approach guaranteed the inclusion of spatial and channel connections throughout the entire feature dimension, maintaining feature recyclability and, as a result, lowering the computational burden of the self-attention mechanism. Most Transformer-based U-Net architectures either combine CNN and Transformer or solely adopt a pure Transformer structure. These approaches either fail to consider the role of CNN in local feature extraction or employ CNN for local feature extraction and Transformer for global feature extraction, performing self-attention operations on features from the CNN network without addressing the computational cost issues of the Transformer's self-attention mechanism. While they manage to maintain a certain level of local and global feature modeling, they often incur high computational costs and model parameters. Balancing network performance and size becomes a challenging endeavor. In light of these considerations, we endeavor to construct an efficient CNN-Transformer U-shaped network.

## III. PROPOSED METHOD

Fig. 1 illustrates the complete structure of the LUCF-Net, which employs an asymmetrical CNN-Transformer U-shaped framework. The central component is the local-global feature extraction module (LG Block), seamlessly integrated within the down-sampling structure of the encoder. Details about each component is described in subsequent sections.

### A. Local-Global Feature Extraction

The global modeling capability of a model in medical image processing plays a pivotal role in the overall feature extraction capability. Researchers have confirmed the significant impact of self-attention on addressing global background in images or long-range spatial dependencies [33]. However, self-attention has to deal with considerable spatial redundancy within the images, such as semantically similar features in nearby regions [34]. Consequently, even considering all tokens on down-sampled feature maps can lead to inefficiency, squandering substantial computational resources. To mitigate this challenge while retaining both global and local contextual information, EdgeViTs introduced
a novel approach to tackle the issue. In contrast to conventional transformer blocks that execute self-attention for every spatial position, its self-attention module focuses on computing self-attention solely for a subset of tokens. Despite this, it manages to attain a comprehensive spatial interaction, much like the standard Multi-Head Self-Attention. Motivated by this approach, we seamlessly incorporated sparse self-attention into a U-shaped network. This integration enables us to enhance our model's ability for local-global modeling while simultaneously lowering computational demands.

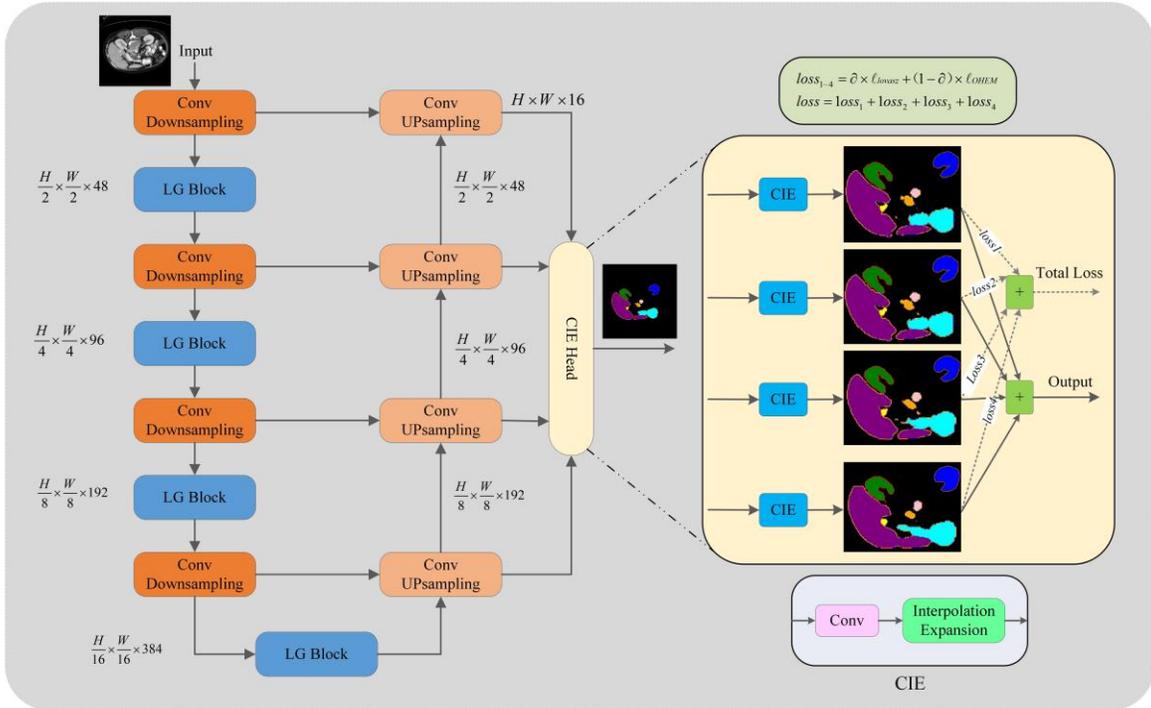

Fig. 1. General layout of the proposed LUCF-Net which involves the adoption of multi-output cascade aggregation and the incorporation of multiple layers of loss during training. LUCF-Net consists of multiple layers of CNN-Transformer encoders and CNN decoders; the LG block does not change the feature map size; the multi-layer outputs are unified into a common format by the CIE head; the training loss is computed as the cumulative sum of the losses associated with the four-layer features located at the decoding portion; the ultimate output of the network results from the aggregation of features derived from all four layers of the decoding portion.

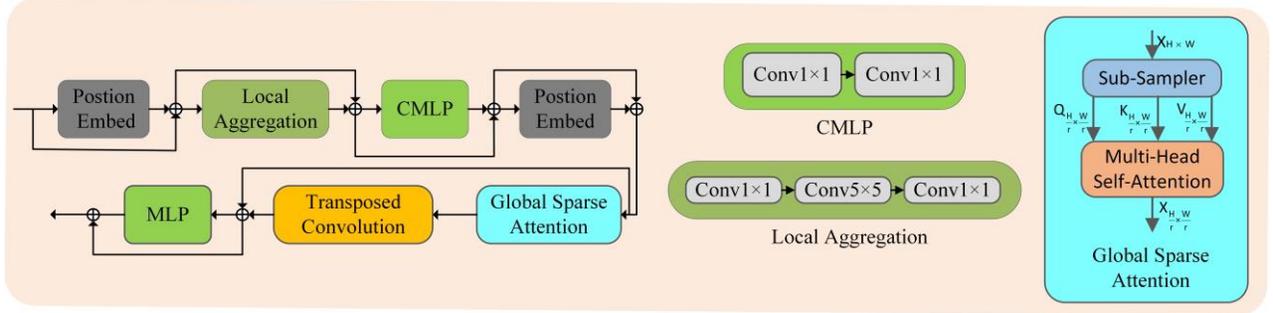

Fig. 2. Structure of the proposed LG Block. LocalAggregation represents the local feature aggregation operator, and Local feature aggregation is realized by multi-layer CNN network; Conv MLP (CMLP) denotes a double-layer perceptron composed of two convolutional layers; Global Sparse Attention is self-attention after sampling operation; Transposed Convolution restores the feature map resolution; MLP is a perceptron consisting of two fully connected layers.

To achieve this, we introduce a local-global feature extraction module called LG Block, as shown in Fig. 2. It takes incoming feature information and initiates a local feature aggregation operation to converge information into local windows. Subsequently, attention is performed on tokens obtained through uniform window sampling. Finally, the globally contextual information obtained from attention operations is propagated through neighborhood spreading using transposed convolutions. The formulation of this module can be described as follows:

$$x = LocalAggregation(x_{in}) + x_{in} \tag{1}$$

$$y = CMLP(x) + x \tag{2}$$

$$z = TransConv(GlobalSparseAttention(y)) + y \tag{3}$$

$$x_{out} = MLP(z) + z \tag{4}$$

## B. Encoder and Decoder

In the encoder section, the initial input image experiences two convolutional layers, which are subsequently followed by down-sampling. After this process, the original input's resolution is reduced by half, while the channel count increases accordingly. Subsequently, the down-sampled image is inputted into the LG Block, where self-attention operations are performed. This sequence is repeated for four layers of convolutional down-sampling and four LG Blocks. Only convolution and up-sampling operations are used in the decoding layer. Similarly, the image resolution is doubled for each layer of convolution and up-sampling. This process is repeated through four up-sampling layers, producing an image size that matches the original input. It's important to highlight that the decoder part does not use the LG Blocks, and the information at both the local and global levels from the encoder can be fused by using skip links and multi-layer cascade modules, thereby avoiding the use of Transformer in the decoder and reducing the model size.

## C. Feature Fusion

Traditional U-Net networks typically utilize the final layer of the decoding side as the comprehensive network output and compute the loss during training. To improve segmentation in the case of multi-scale image outputs, the architecture is enhanced by integrating multi-scale information from different encoder layers into the decoder through skip connections. Each layer's up-sampled output from the decoder is fed into a separate decoupled head. This decoupled head corresponds to the CIE head in Fig. 1, which coordinates images of different scales into consistent output sizes. The CIE head uses a bilinear interpolation operation, which is the same as the upsampling operation on the Decoder. During training, we compute the loss by comparing each layer's output accompanied by its corresponding label. The final output is a summation of outputs from four distinct stages. This structure intensifies spatial relationships between pixels through multi-level cascading, expediting model convergence during training.

## D. Loss Function

In medical image segmentation, the cross-entropy loss and Dice loss stand out as the most commonly favored loss functions [25]. The Dice coefficient is a commonly employed metric in the visual computing domain, used to measure the similarity between two images. However, the training of the Dice loss exhibits significant fluctuations, therefore, it is frequently integrated with the cross-entropy loss function. Here, we introduce the Lovász Softmax loss [35] as a replacement for the Dice loss. We made this choice because this loss also directly optimizes the region-based metrics. It is a convex function, ensuring that it does not get stuck in local minima during training. Moreover, the Lovász Softmax loss performs well in handling object boundary pixels, avoiding the generation of blurry edges. The Lovász Softmax loss originates from a variant of the Jaccard index loss, which of category $c$ can be represented using the following formula:

$$\Delta J_c = 1 - \frac{\sum_{i=1}^{M}(y_i^c \cap p_i^c)}{\sum_{i=1}^{M}(y_i^c \cup p_i^c)} \quad (5)$$

Here, $y_i^c$ and $p_i^c$ represent the label and network's prediction for the $i$-th pixel, respectively. $c$ is a subclass of $C$ which denotes the total number of categories, and $M$ is the number of pixels for a batch. Formula (5) is a discrete function, which is not suitable for direct optimization of loss. The Lovász extension is employed to confer differentiability upon the Jaccard index loss, thereby converting discrete input values into continuous values. After adopting the extension, the Lovász Softmax loss function can be computed through the subsequent equations:

$$e_i(c) = \begin{cases} 1 - f_i(p_i^c) & \text{if } c = y_i^c \\ f_i(p_i^c) & \text{otherwise} \end{cases} \quad (6)$$

$$\ell_{Lovasz} = \frac{1}{|C|} \sum_{c \in C} \overline{\Delta J_c}(e(c)), \quad (7)$$

Here, $f_i(p_i^c)$ is the probability distribution of the network output on category $c$, which is obtained by the Softmax function. $e_i(c)$ is the pixel error of category $c$, and the vector $e(c)$ is the substitute of the Jaccard index of category $c$. $\overline{\Delta J_c}$ is the Lovász extension of Jaccard index.

To mitigate the problem of sample imbalance within the dataset, the Online Hard Example Mining (OHEM) loss function [36] is introduced. During the training of deep learning models, this loss function strategy is utilized to tackle problems arising from imbalances in class distribution. The purpose of the OHEM loss is to focus on challenging-to-classify samples, encouraging the model to better learn difficult cases and thereby enhance overall performance. Throughout the training phase, the core idea of OHEM loss is to select difficult-to-classify samples from the batch for backpropagation. This effectively directs the model's attention towards challenging instances, aiding the model in better distinguishing between different categories. Naturally, we extend the definition of hard samples in the OHEM loss function to the pixel level. For each set of training batch, the initial loss function calculates the average cross-entropy loss for all pixels trained in the current batch. The cross-entropy loss is expressed by:

$$CE = -\sum_{c=1}^{C} y_i^c \log(f_i(p_i^c)) \quad (8)$$

Based on cross-entropy loss, the OHEM loss formula can be articulated as:

$$\ell_{org} = \frac{1}{M} \sum_{i=1}^{M} CE_i, \quad (9)$$

$$\ell_{re} = \frac{1}{K} \sum_{i=1}^{K} CE_i, \quad (10)$$

$$\ell_{OHEM} = \ell_{org} + \ell_{re} \quad (11)$$

Here, $\ell_{org}$ and $\ell_{re}$ represent the loss for all pixels and the loss associated with the selected hard pixels, respectively. The variable $K$ represents the number of hard pixels, which are determined by filtering out prediction pixels with low confidence. The OHEM loss selects these pixels with low confidence and computing the mean cross-entropy loss. Subsequently, the average loss of hard pixels is aggregated with the average cross-entropy loss of all pixels.

In summary, our hybrid loss can be determined as follows:

$$loss = \partial \times \ell_{lovasz} + (1 - \partial) \times \ell_{OHEM} \quad (12)$$

IV. EXPERIMENTS

*A. Datasets and Evaluation Metrics*

To assess the network's generalization performance, experimental tests were carried out using three distinct dataset categories. The Synapse multi-organ abdominal dataset consists of CT format data, the Automated Cardiac Diagnosis Challenge (ACDC) dataset comprises data in the MRI format, and the ISIC2016 and ISIC2018 datasets are composed of image format data. With the exception of ISIC2018, which employed a five-fold cross-validation approach, the remaining experimental outcomes were derived from the mean and standard deviation of five experiments.

*1) Synapse Dataset:* Synapse abdominal multi-organ dataset [37] comprises 30 CT scans of the abdominal region and 3779 abdominal clinical CT images captured in the axial plane. The dataset has been partitioned into 18 training scans and 12 random test scans. We used the same processing method according to [27]. We took the last round of training results as the test weight. The evaluation metrics are the mean DSC and mean HD for 8 abdominal organs.

*2) ACDC Dataset:* Automated cardiac diagnosis challenge (ACDC) dataset [38]: The dataset comprises 100 MRI scans obtained from various patients, with each scan annotated for three organs: the left ventricle (LV), right ventricle (RV), and myocardium (MYO). Following the [43], 70 cases were allocated for training, 10 for validation, and 20 for testing. The average DSC served as the performance metric for evaluation.

*3) ISIC Datasets:* For ISIC-2016 [39] dataset, there are 900 training samples and 379 validation samples in total. The ISIC-2018 [40] dataset comprises 2594 images along with their corresponding labels, and the image resolutions range from 720×540 to 6708×4439. As stated in reference [41], a five-fold cross-validation was conducted to guarantee an equitable evaluation. The assessment was based on the average DSC and average Intersection over Union (IoU) scores.

## B. Implementation Details

The experiments were conducted utilizing PyTorch 2.0.0 framework, with training executed on an Nvidia RTX 3090 GPU boasting 24 GB of memory. For Synapse dataset and ACDC dataset, the specified input image dimensions were configured as 224×224. There was only one channel, and the batch size used during the training was 16. For the two ISIC datasets, the input image dimensions were configured as 512×512, employing 3 channels, and a batch size of 4 was employed during training. We have adopted a dynamic learning rate for training. Specifically, the initial rate of learning is 0.05. During the training process, the learning rate gradually decreases as the number of training epochs increases. LUCF-Net underwent fine-tuning with the SGD optimizer, with a momentum of 0.9 and a weight decay of 0.0001. During the experiments, data augmentation methods like flipping and rotation were used to increase the variety of the data. In addition, we used the matplotlib function when drawing the result image [42].

## C. Experimental Results

*1) Results on Synapse Dataset:* Comparison between our proposed LUCF-Net and SOTA methods is presented on multi-organ dataset of the abdomen as shown in Table I. The last two columns represent the average Dice Similarity Coefficient (DSC) and the average Hausdorff Distance (HD) of the 8 organs. The values below for different organs represent the average DSC. Compared with other models based on CNN or Transformer, LUCF-Net exhibits a 1.54% lead over the TransCASCADE [43] in terms of DSC, and a 2.60 millimeters (mm) lead in terms of HD. The figures in Fig. 3 depict the segmentation outcomes achieved by various methods on the multi-organ CT dataset. These images demonstrate that LUCF-Net accurately delineates complex structures in most organ segmentation tasks, yielding more precise segmentation results and exhibiting competitive performance even in challenging backgrounds.

TABLE I
PERFORMANCE COMPARISONS OF VARIOUS TECHNIQUES ON SYNAPSE ABDOMINAL MULTI-ORGAN DATASET
(OVERALL DICE SCORE% AND HAUSDORFF DISTANCE IN MM, AND DICE SCORE % FOR EACH ORGAN)

| Methods | Stomach | Spleen | Pancreas | Liver | Kidney(R) | Kidney(L) | Gallbladder | Aorta | DSC↑ | HD↓ |
|---|---|---|---|---|---|---|---|---|---|---|
| U-Net [4] | 56.98 | 86.67 | 53.98 | 93.43 | 68.60 | 77.77 | 69.72 | 89.07 | 76.85 | 39.70 |
| DARR [44] | 45.96 | 89.90 | 54.18 | 94.08 | 73.24 | 72.31 | 53.77 | 74.74 | 69.77 | - |
| V-Net [25] | 75.58 | 80.56 | 40.05 | 87.84 | 80.75 | 77.10 | 51.87 | 75.34 | 68.81 | - |
| R50 Att-Unet [27] | 74.95 | 87.19 | 49.37 | 93.56 | 72.71 | 79.20 | 63.91 | 55.92 | 75.57 | 36.97 |
| TransUnet [27] | 75.62 | 85.08 | 55.86 | 94.08 | 77.02 | 81.87 | 63.13 | 87.23 | 77.48 | 31.69 |
| Swin-Unet [28] | 76.60 | 90.66 | 56.58 | 94.29 | 79.61 | 83.28 | 66.53 | 85.47 | 79.13 | 21.55 |
| MISSFormer [31] | 80.81 | 91.92 | 65.67 | 94.41 | 82.00 | 85.21 | 68.65 | 86.99 | 81.96 | 18.20 |
| DAEFormer [32] | 79.19 | **91.94** | 65.12 | 94.98 | 80.88 | 86.08 | **72.30** | 88.96 | 82.43 | 17.46 |
| HiFormer [41] | 82.03 | 90.44 | 60.77 | 94.07 | 78.37 | 84.23 | 68.61 | 87.03 | 80.69 | 19.14 |
| PVT-CASCADE [43] | 83.69 | 90.01 | 64.43 | 94.08 | 80.37 | 82.23 | 70.59 | 83.01 | 81.06 | 20.23 |
| TransCASCADE [43] | 83.52 | 90.79 | 65.33 | 94.43 | 84.56 | 87.66 | 68.48 | 86.63 | 82.68 | 17.34 |
| LUCF-Net | **85.05** ±0.74 | 91.54 ±0.80 | **67.32** ±1.29 | **95.52** ±0.15 | **85.04** ±0.69 | **87.88** ±0.67 | 71.14 ±1.40 | **89.66** ±0.23 | **84.22** ±0.31 | **14.74** ±1.14 |

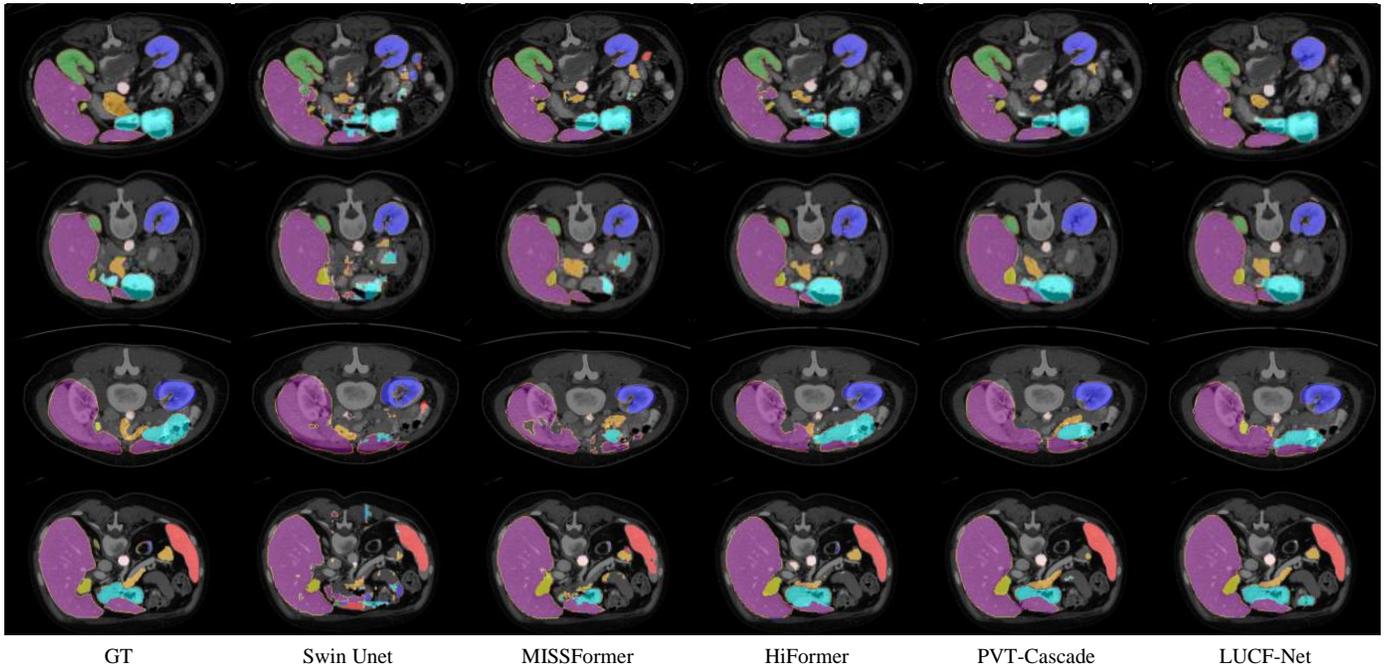

| GT | Swin Unet | MISSFormer | HiFormer | PVT-Cascade | LUCF-Net |

Fig. 3. Comparison of various techniques on the Synapse abdominal multi-organ dataset through visual analysis. Each row represents a different subject, the first column is the true label for each subject, and different colors represent different organs.

*2) Results on ACDC Dataset:* Table II displays the comparison of the performance between LUCF-Net and SOTA methods on the ACDC dataset. The final column represents the average DSC across three segments of the heart, while the initial three columns display the average DSC for distinct segments. Notably, LUCF-Net has attained the highest average DSC at 92.19%, as illustrated in Fig. 4.

TABLE II
PERFORMANCE COMPARISONS OF VARIOUS TECHNIQUES ON AUTOMATED CARDIAC DIAGNOSIS CHALLENGE DATASET
(OVERALL DICE SCORE %, AND DICE SCORE % FOR EACH LOCATION)

| Methods | Myo | LV | RV | DSC↑ |
|---|---|---|---|---|
| R50 U-Net [27] | 80.63 | 94.92 | 87.10 | 87.55 |
| R50 ViT [27] | 81.88 | 94.75 | 86.07 | 87.57 |
| TransUnet [27] | 84.53 | 95.73 | 88.86 | 89.71 |
| Swin Unet [28] | 85.62 | 95.83 | 88.55 | 90.00 |
| MISSFormer [31] | 88.04 | 94.99 | 89.55 | 90.86 |
| TransCASCADE [43] | **90.25** | 95.50 | 89.14 | 91.63 |
| PVT-CASCADE [43] | 89.97 | 95.50 | 88.9 | 91.46 |
| LUCF-Net | 90.13 ±0.10 | **95.96** ±0.08 | **90.46** ±0.16 | **92.19** ±0.08 |

*3) Results on Skin Lesion Segmentation:* Table III shows the comparison results of the average DSC and average IoU performance of LUCF-Net and other networks on ISIC 2016 and ISIC 2018. It is noticeable that LUCF-Net also demonstrated competitive performance in processing image datasets. Visual comparisons of the results are also depicted in the Fig. 5. Results support that LUCF-Net is capable of capturing intricate details and producing more precise contours. In contrast to a pure Transformer network, our approach captures finer local details, indicating its effectiveness in extracting local and global features.

TABLE III
PERFORMANCE COMPARISONS OF VARIOUS TECHNIQUES ON BOTH ISIC2016 AND ISIC2018 LESION SEGMENTATION
(DICE SCORE % AND IoU SCORE %)

| Method | ISIC2016 | | ISIC2018 | |
|---|---|---|---|---|
| | DSC↑ | IoU↑ | DSC↑ | IoU↑ |
| UNet [4] | 89.84 | 83.15 | 83.86 | 75.07 |
| UNet++ [18] | 88.89 | 82.12 | 84.09 | 75.26 |
| UNet3+ [20] | 89.09 | 82.43 | 83.05 | 74.00 |
| UCTransNet [45] | 90.58 | 84.14 | 86.71 | 78.43 |
| UNeXt-L [46] | 90.50 | 84.36 | 86.72 | 78.91 |
| STM-Unet [47] | 90.94 | 84.63 | 87.51 | 79.84 |
| LUCF-Net | **91.18** ±0.10 | **84.98** ±0.44 | **89.85** ±0.46 | **83.32** ±0.58 |

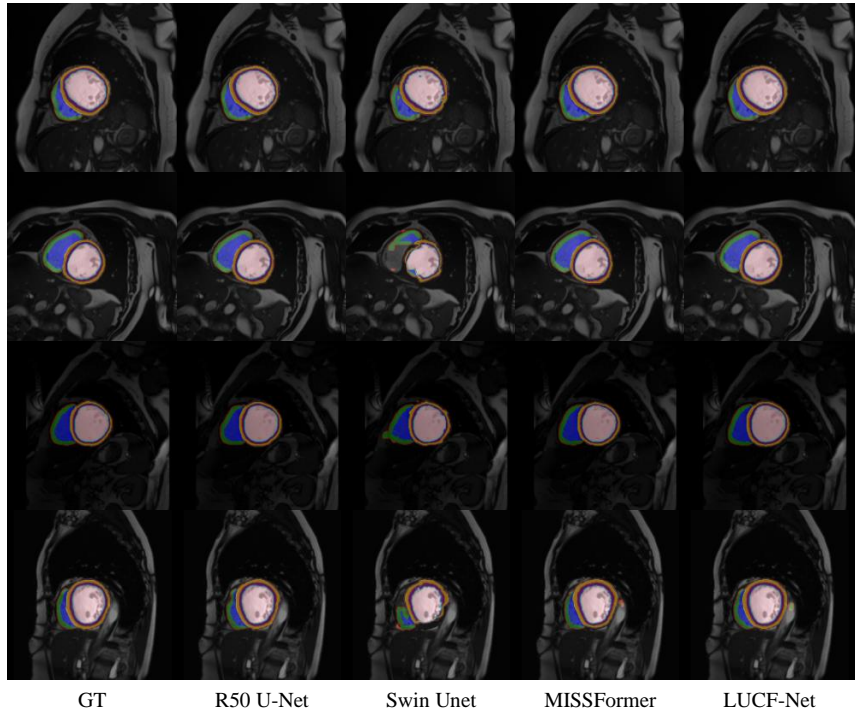

Fig. 4. Comparison of various techniques on the ACDC dataset through visual analysis. Each row represents a different subject, the first column is the true label for each subject, and the different colors represent the different structures of the heart.

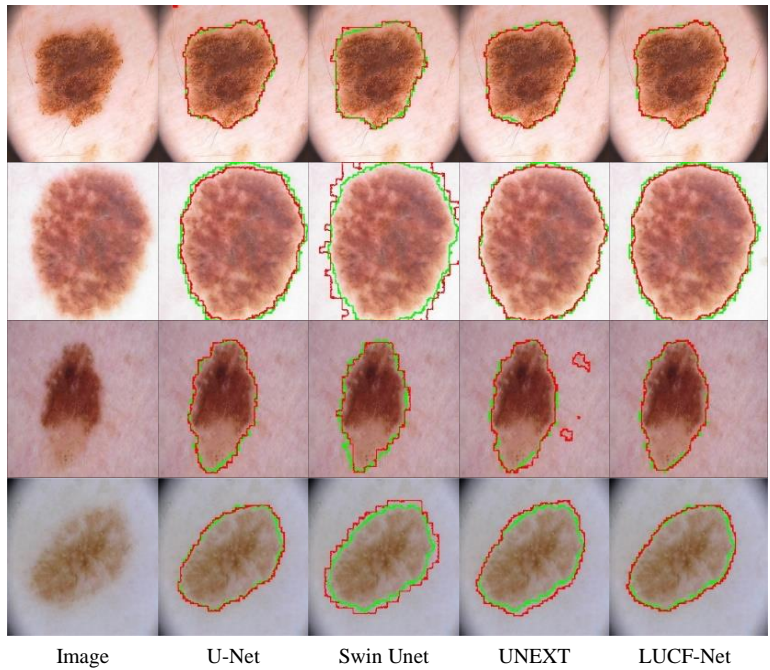

Fig. 5. Comparison of various techniques on the ISIC dataset through visual analysis. Each row represents a different sample. The first column is a picture of a different sample. The dataset has two categories, only foreground and background. The green outline represents the range of the real label, and the red label represents the segmentation range predicted by the network.

*4) Attention Layer Analysis:* To further validate the effectiveness of modeling local and global features in the network, we conducted a study from the perspective of the attention layer. We chose the state-of-the-art U-shaped network architecture, which shares similarities in downsampling encoding design with Swin Unet [28], MISSFormer [31], and LUCF-Net. We selected four layers of feature maps processed through self-attention layers for comparison. In the case of DAEFormer [32], it has only three layers of downsampling in the encoding stage, and we selected the feature maps of these three layers processed through sub-attention layers for comparison. As shown in Fig.6, compared to other networks, LUCF-Net captures more detailed image information in the shallow encoding stage, enabling it to capture finer local feature details. In the deep encoding stage, LUCF-

Net achieves better global modeling of target information, resulting in more complete representations of the objects contained in the feature information.

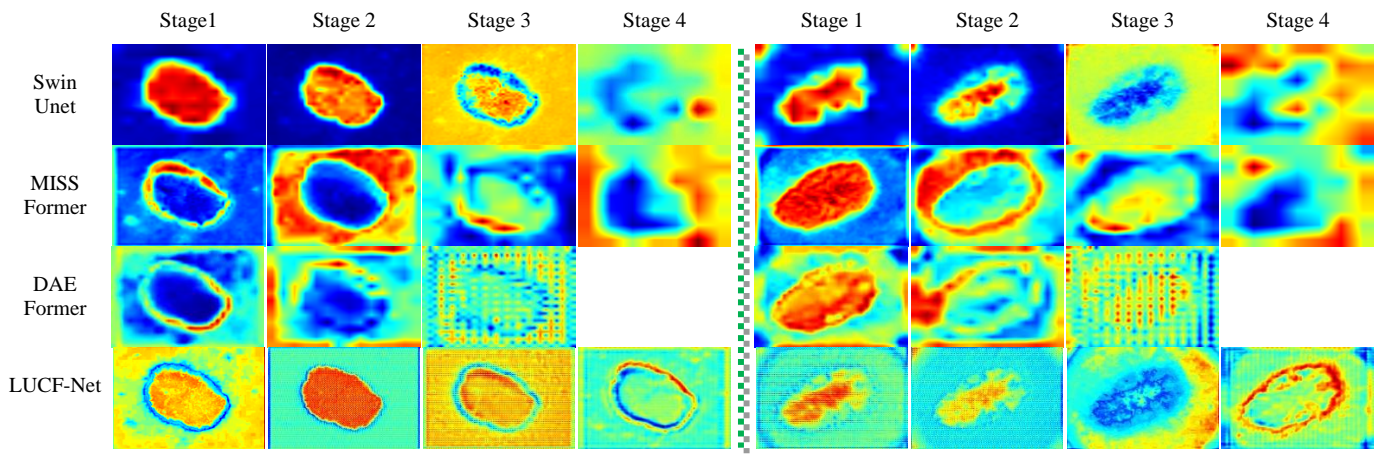

Fig. 6. Comparison of attention layer feature visualization. Each section separated by dashed lines represents the visualized sampling features at different stages of different networks on the same image. DAEFormer has three stages of downsampling, while other networks have four stages of downsampling.

*5) Comparison of Model Parameters:* In Table IV, we've compared our proposed method against the parameter counts of medical image segmentation models for comparison. The input shape of the network is standardized as 1×1×224×224 and quantified the computational intensity of the model using Params and GFLOPs. Params represents the number of parameters in a neural network, while GFLOPs indicates the order of magnitude of the number of floating-point operations per second performed by the model during inference or training. Unlike the SOTA Transformer-based networks, our model demonstrates a certain advantage in terms of complexity. Specifically, we achieve performance beyond existing high-complexity Transformer networks while leveraging the simplicity of the CNN architecture.

TABLE IV
COMPLEXITY COMPARISONS OF DIFFERENT METHODS

| Methods | Params (M) | GFLOPs | DSC↑ |
|---|---|---|---|
| R50 U-Net [27] | 13.04 | 61.90 | 75.57 |
| TransUNet [27] | 96.07 | 88.91 | 77.48 |
| SwinUnet [28] | 29.58 | **5.25** | 79.13 |
| MISSFormer [31] | 35.45 | 7.26 | 81.96 |
| DAEFormer [32] | 29.69 | 26.16 | 82.43 |
| HiFormer [41] | 34.14 | 17.81 | 80.69 |
| PVT-CASCADE [43] | 35.27 | 6.24 | 81.06 |
| LUCF-Net | **6.93** | 6.60 | **84.22±0.31** |

## D. Ablation Study

We first analyzed the effectiveness of the proposed framework from two aspects: network architecture and loss functions. Then, we examined the impact of hyper-loss parameters on model performance. All experiments were conducted on the Synapse dataset, using DSC and HD as evaluation metrics. The details of the ablation experiments are presented below.

*1) Ablation Experiment on Network Architecture:* To assess the efficacy of the network architecture, we employed a dual loss function comprising cross-entropy and Dice loss during the ablation study on the network structure. The hyperparameter configuration followed the guidelines outlined in [32], as detailed in Table V. Incorporating both local and global feature extraction modules, either individually or in combination for feature fusion, led to notable enhancements in network performance. Simultaneous utilization of both modules yielded a 1.24% increase in DSC and a 6.54 mm increment in HD, surpassing the outcomes attained through the use of individual modules.

TABLE V
PERFORMANCE OF DIFFERENT
MODULE COMBINATIONS OF LUCF-NET

| LG | Fusion | DSC↑ | HD↓ |
|---|---|---|---|
| × | × | 81.82±0.69 | 22.97±0.98 |
| × | √ | 81.91±0.68 | 22.61±3.32 |
| √ | × | 82.82±0.31 | 19.33±1.58 |
| √ | √ | **83.06±0.45** | **16.43±1.45** |

*2) Ablation Experiment on Loss Function:* To assess the effectiveness of the loss combination used, we conducted experiments using different loss functions while training LUCF-Net under consistent experimental conditions. Given that loss computation involves multi-level feature fusion, we integrated feature fusion into the ablation study of loss function variables. The OHEM loss originates from cross-entropy loss, while the Lovász Softmax loss and Dice loss quantify set similarity. We replaced cross-entropy loss and Dice loss with OHEM loss and Lovász Softmax loss, respectively. As shown in Table VI, the combination of Lovász Softmax loss and OHEM loss has been proven more suitable for feature fusion networks. Compared to the combination of cross-entropy loss and Dice loss, DSC increases by 1.16% and HD decreases by 1.69 mm.

TABLE VI
ABLATION EXPERIMENT ON LOSS FUNCTION

| OHEM loss | Lovász Softmax loss | Fusion | DSC↑ | HD95↓ |
|---|---|---|---|---|
| × | × | × | 82.82±0.31 | 19.33±1.58 |
| √ | × | × | 82.96±0.84 | 19.14±2.68 |
| × | √ | × | 82.81±0.60 | 19.88±2.38 |
| √ | √ | × | 82.80±0.69 | 18.04±2.63 |
| × | × | √ | 83.06±0.45 | 16.43±1.45 |
| √ | × | √ | 83.48±0.65 | 17.90±2.45 |
| × | √ | √ | 83.64±0.45 | 14.78±2.05 |
| √ | √ | √ | **84.22±0.31** | **14.74±1.14** |

Furthermore, in Fig. 7, the training loss curves for various combinations of loss functions on the proposed final network architecture were depicted. Networks trained with OHEM loss and Lovász Softmax loss demonstrate enhanced stability compared to those trained with other loss functions. Expanding upon the utilization of OHEM loss and Lovász Softmax loss, Table VII outlines the effects of fusing different layers on model performance, while Fig. 8 illustrates the training loss curves for different layers. The four hierarchical features input into the CIE Head are derived from the four feature maps output by the decoder, all of which undergo LeakyReLU operations. The determination of CIE Head values relies solely on convolutional and bilinear interpolation techniques, ensuring uniformity in output shape. We separately took the sums of outputs from the last layer, the last two layers, the last three layers, and the last four layers as the results for fusing features from different layer depths. The results indicate that as fusion layers deepen, there is noticeable improvement in both model performance and network convergence speed.

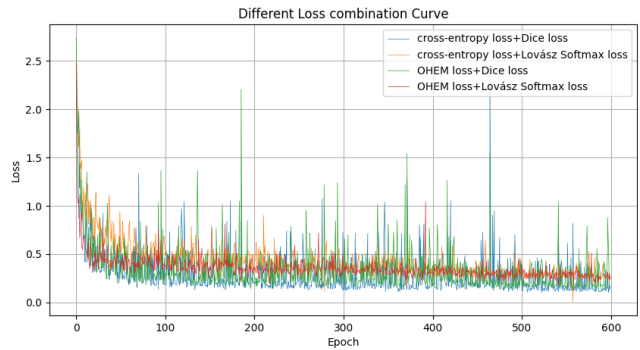

Fig. 7. Loss curves for different combinations of loss functions.

TABLE VII
ABLATION EXPERIMENT ON FUSION STRUCTURE

| Number of Layers | DSC↑ | HD↓ |
|---|---|---|
| 1 | 82.80±0.69 | 18.04±2.63 |
| 2 | 82.91±0.42 | 15.49±3.31 |
| 3 | 83.33±0.17 | 15.53±0.35 |
| 4 | **84.22±0.31** | **14.74±1.14** |

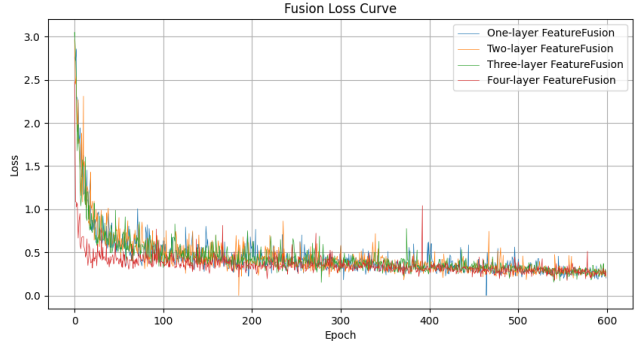

Fig. 8. Loss curves of different fusion levels.

*3) Ablation Experiment on Loss Hyperparameters:* We employed a grid search approach to explore the impact of different hyperparameter combinations of loss functions on model performance, as illustrated in Fig. 9. The performance of models trained solely with Lovász Softmax loss or OHEM loss is suboptimal, especially when using only OHEM loss as the training loss function. OHEM loss prioritizes pixel-level classification accuracy, while Lovász Softmax loss focuses on measuring the similarity between two sets. Through the optimized combination, better results can be achieved.

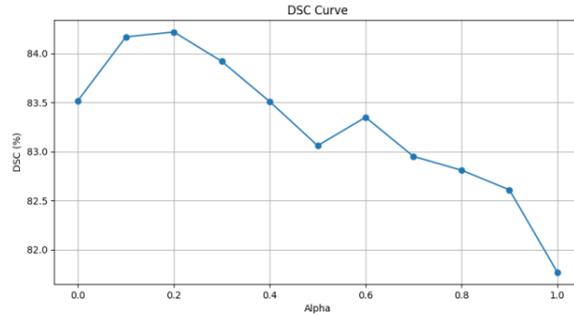

Fig. 9. Model performance with different weights of loss functions.

## V. Discussion

The current approaches that merge CNN and Transformer typically demand significant computational and storage resources. To address these issues, we have incorporated a sparse self-attention Transformer module, integrated sparse self-attention into each encoder level, and fused CNN layers with Transformer to efficiently capture both local and global features. When comparing LUCF-Net's performance on four datasets in CT, MRI, and picture formats, it consistently outperformed the current SOTA approaches. The visualized results demonstrated that LUCF-Net effectively captures both local details and overarching global data, showcasing its strong performance. In addition, the complexity of LUCF-Net outperformed other networks owing to its intricate design, which features a simplistic Transformer module and an asymmetric network architecture.

Due to its low complexity and outstanding segmentation performance, LUCF-Net has the potential to serve as a dependable backbone network. While existing methods such as HiFormer [41] and PVT-CASCADE [43] integrate CNN and Transformer, they do not adequately solve the underlying issue of the computational burden caused by the self-attention mechanism. Efficient pure Transformer U-shaped networks like MISSFormer [31] may disregard the advantages of CNN networks in regional feature extraction, while the symmetrical Transformer module design also adds complexity to the network. The proposed LUCF-Net in this study effectively reduces the computational requirements of the Transformer module by utilizing an efficient and sparse self-attention mechanism. It also incorporates an asymmetric design and a multi-layer feature cascade fusion mechanism, simplifying the network while improving the CNN's ability to extract local features. As a result, LUCF-Net demonstrates strong performance in medical image segmentation tasks.

## VI. Conclusion

This study introduced a novel medical image segmentation approach, termed LUCF-Net, through the combination of CNN and Transformer. In contrast to other SOTA models based on CNN and Transformer, the LUCF-Net not only captures more detailed image information but also achieves better global modeling of target information. Modeling these global features helps the network to better comprehend the entire image context, thereby enhancing segmentation performance. LUCF-Net also showed better segmentation performance with reduced model complexity, showing the potential for medical image segmentation applications. However, in medical image processing, there is often a challenge of insufficient data due to the relatively limited number of available samples, making it difficult to support fully supervised training. To address this challenge, Wang et al. [48]

have employed an innovative approach by combining a visual Transformer with a consistency regularization framework to achieve superior performance compared to semi-supervised segmentation frameworks based on CNN networks with limited annotated data. In future work, our approach can be integrated with semi-supervised medical image segmentation to further enhance segmentation performance.

## References

[1] W. Shen, G. Wu, and H. I. Suk, "Deep learning in medical image analysis," *Aneu Rev Biomed Eng.*, 2017, vol. 19, pp. 221-248.
[2] D. D. Patil and S. G. Deore, "Medical image segmentation: A review," *Int. J. Comput. Sci. Mobile Comput*., 2013, vol. 2, no. 1, pp. 22-27.
[3] A. Vaswani et al., "Attention is all you need," *Adv. Neural Inf. Process., Syst*., 2017, vol. 30.
[4] O. Ronneberger, P. Fischer, and T. Brox, "U-Net: Convolutional networks for biomedical image segmentation," *in Proc. Int. Conf. Med. Image Comput. Comput.- Assist. Intervention.*, 2015, pp. 234–241.
[5] A. W. Salehi, S. Khan, G. Gupta, et al., "A study of CNN and transfer learning in medical imaging: advantages, challenges, future scope," *Sustainability-Basel.*, 2023, vol. 15, no. 7, pp. 5930.
[6] C. You, R. Zhao, F. Liu, et al., "Class-aware adversarial transformers for medical image segmentation," *Adv. Neural Inf. Process. Syst.*, 2022, vol. 35, pp. 29582-29596.
[7] M. M. Naseer, K. Ranasinghe, S. H. Khan, et al., "Intriguing properties of vision transformers," *Adv. Neural Inf. Process. Syst*., 2021, vol. 34, pp. 23296-23308.
[8] E. Xie, W. Wang, Z. Yu, et al., "SegFormer: Simple and efficient design for semantic segmentation with transformers," *Adv. Neural Inf. Process. Syst*., 2021, vol. 34, pp. 12077-12090.
[9] F. Yuan, Z. Zhang, Z. Fang, "An effective CNN and transformer complementary network for medical image segmentation," *Pattern Recogn*., 2023, vol. 136, pp. 109228.
[10] S. Khan, M. Naseer, M. Hayat, et al., "Transformers in vision: A survey," *ACM Comput. Surv.*, 2022, vol. 54, no. 10, pp. 1-41.
[11] J. Pan, A. Bulat, F. Tan, et al., "Edgevits: Competing light-weight CNNs on mobile devices with vision transformers," *in Proc. Eur. Conf. Comput. Vis. Pattern Recogn.*, 2022, pp. 294-311.
[12] S. Nahian, P. Sidike, C. P. E. Colin, V. D., et al., "U-Net and its variants for medical image segmentation: A review of theory and applications," *IEEE Access.*, 2021, vol. 9, pp. 82031-82057.
[13] F. I. Diakogiannis, F. Waldner, P. Caccetta, et al., "ResUNet-a: A deep learning framework for semantic segmentation of remotely sensed data," *ISPRS J Photogramm.*, 2020, vol. 162, pp. 94-114.
[14] K. He, X. Zhang, S. Ren, J. Sun, et al., "Deep residual learning for image recognition," *in Proc. IEEE Conf. Comput. Vis. Pattern Recogn*., 2016, pp. 770-778.
[15] E. Thomas, S. J. Pawan, S. Kumar, et al. Multi-res-attention UNet: a CNN model for the segmentation of focal cortical dysplasia lesions from magnetic resonance images[J]. IEEE Journal of Biomedical and Health Informatics, 2020, 25(5): 1724-1734.
[16] N. T. Do, S. T. Jung, H. J. Yang, et al. Multi-level seg-unet model with global and patch-based X-ray images for knee bone tumor detection[J]. Diagnostics, 2021, 11: 691.
[17] S. Guan, A. A. Khan, S. Sikdar, et al., "Fully dense UNet for 2-D sparse photoacoustic tomography artifact removal," *IEEE J. Biomed.Health Inform.*, 2019, vol. 24, no. 2, pp. 568-576.
[18] Z. Zhou, M. M. R. Siddiquee, N. Tajbakhsh, and J. Liang, "UNet++: Redesigning skip connections to exploit multiscale features in image segmentation," *IEEE Trans. Med. Imag*., 2020, vol. 39, no. 6, pp. 1856–1867.
[19] J. Lin, Q. She, Y. Chen, "Pulmonary nodule detection based on IR-UNet + +," *Med. Biol. Eng. Comput*., 2023, vol. 61, no. 2, pp. 485-495.
[20] H. Huang, L. Lin, R. Tong, et al., "Unet 3+: A full-scale connected UNet for medical image segmentation," *in Proc. ICASSP 2020-2020 IEEE Int. Conf. Acoust., Speech Signal Process*., 2020, pp. 1055-1059.
[21] S. Cai, Y. Tian, H. Lui, et al., "Dense-UNet: A novel multiphoton in vivo cellular image segmentation model based on a convolutional neural network," *Quant. Imaging Med. Surg*., 2020, vol. 10, no. 6, pp. 1275.
[22] A. R. Feyjie, R. Azad, M. Pedersoli, et al., "Semi-supervised few-shot learning for medical image segmentation," 2020, *arXiv:2003.08462*.
[23] S. Peng, W. Chen, J. Sun, et al., "Multi‐scale 3d u-nets: An approach to automatic segmentation of brain tumor," *Int. J. Imaging Syst. Technol.*, 2020, vol. 30, no. 1, pp. 5-17.
[24] Ö. Çiçek, A. Abdulkadir, S. S. Lienkamp, et al., "3D U-Net: learning dense volumetric segmentation from sparse annotation," *in Proc. Int. Conf. Med. Image Comput. Comput.- Assist. Intervention*., 2016, pp. 424-432.
[25] F. Milletari, N. Navab, S. A. Ahmadi, "V-net: Fully convolutional neural networks for volumetric medical image segmentation," *in Proc. Fourth Int. Conf. 3D Vision.*, 2016, pp. 565-571.
[26] A. Dosovitskiy, L. Beyer, A. Kolesnikov, et al., "An image is worth 16x16 words: Transformers for image recognition at scale," 2020, *arXiv:2010.11929*.
[27] J. Chen et al., "TransuNet: Transformers make strong encoders for medical image segmentation," 2021, *arXiv:2102.04306*.
[28] H. Cao et al., "Swin-Unet: Unet-like pure transformer for medical image segmentation," *in Proc. Eur. Conf. Comput. Vis*., 2022, pp. 205–218.
[29] Z. Liu, Y. Lin, Y. Cao, et al., "Swin transformer: Hierarchical vision transformer using shifted windows," *in Proc. IEEE/CVF Int. Conf. Comput. Vis*., 2021, pp. 10012-10022.
[30] A. Liu, B. Chen, J. Xu, Z. Zheng, G. Liu, et al., "DS-TransUNet: Dual swin transformer U-Net for medical image segmentation," *IEEE T Instrum Meas*., vol. 71, pp. 1-15, 2022.
[31] X. Huang, Z. Deng, D. Li, X. Yuan and Y. Fu, "MISSFormer: An effective transformer for 2D medical image segmentation," *IEEE Trans. Med. Imag.*, vol. 42, no. 5, pp. 1484-1494, 2023.
[32] M. M. Rahman, R. A. René, A. K. A. Ehsan, A. K. Amirhossein, M. Dorit, et al., "DAE-Former: Dual attention-guided efficient transformer for medical Image segmentation," 2023, *arXiv:2212.13504*.
[33] M. Lan, J. Zhang, Z. Wang, "Coherence-aware context aggregator for fast video object segmentation, " *Pattern Recogn*., 2023, vol. 136, pp. 109214.
[34] J. Yang, C. Lin, P. Zhang, X. Duan, B. Xu, Y. Yan, J. Guan, et al., "Focal self-attention for local-global interactions in vision transformers," *CoRR*, 2021, abs/2107.00641.
[35] M. Berman, A. R. Triki, M. B. Blaschko, "The lovász-softmax loss: A tractable surrogate for the optimization of the intersection-over-union measure in neural networks," *in Proc. IEEE Conf. Comput. Vis. Pattern Recogn*., 2018, pp. 4413-4421.
[36] A. Shrivastava, A. Gupta, R. Girshick, "Training region-based object detectors with online hard example mining," *in Proc. IEEE Conf. Comput. Vis. Pattern Recogn*., 2016, pp. 761-769.
[37] MICCAI 2015 Multi-Atlas Abdomen Labeling Challenge, "Synapse multi-organ segmentation dataset," https://www.synapse.org/#! Synapse: syn3193805/wiki/217789, 2015. Accessed: 2022-04-20.

[38] O. Bernard, A. Lalande, C. Zotti, et al., "Deep learning techniques for automatic MRI cardiac multi-structures segmentation and diagnosis: is the problem solved?" *IEEE Trans. Med. Imag*., vol. 37, no. 11, pp. 2514-2525, 2018.

[39] G. David, C. F. C. Noel, C. M. E. Celebi, et al., "Skin lesion analysis toward melanoma detection: A challenge at the international symposium on biomedical imaging (isbi) 2016, hosted by the International Skin Imaging Collaboration (ISIC)," *CoRR*., 2016, abs/1605.01397.

[40] N. C. F. Codella, D. Gutman, M. E. Celebi, et al., "Skin lesion analysis toward melanoma detection: A challenge at the 2017 international symposium on biomedical imaging (isbi), hosted by the international skin imaging collaboration (isic)," *in Proc. 15th IEEE Int. Symp. Biomed. Imag*., 2018, pp. 168-172.

[41] M. Heidari, A. Kazerouni, M. Soltany, et al., "Hiformer: Hierarchical multi-scale representations using transformers for medical image segmentation," *in Proc. IEEE/CVF Winter Conf. Appl. Comput. Vis*., 2023, pp. 6202-6212.

[42] Hunter, J. D. (2007). Matplotlib: A 2D graphics environment. Computing in Science & Engineering, 9(3): 90-95.

[43] M. M. Rahman, R. Marculescu, "Medical image segmentation via cascaded attention decoding," *in Proc. IEEE/CVF Winter Conf. Appl. Comput. Vis*., 2023, pp. 6222-6231.

[44] S. Fu, Y. Lu, Y. Wang, et al., "Domain adaptive relational reasoning for 3D multi-organ segmentation," *in Proc. Int. Conf. Med. Image Comput. Comput.- Assist. Intervention*., 2020, pp. 656-666.

[45] Z. Wang, P. Cao, J. Wang, et al., "Uctransnet: Rethinking the skip connections in U-Net from a channel-wise perspective with transformer," *in Proc. AAAI Conf. Artif. Intell*., 2022, vol. 36, no. 3, pp. 2441-2449.

[46] J. M. J. Valanarasu, V. M. Patel, "Unext: MLP-based rapid medical image segmentation network," *in Proc. Int. Conf. Med. Image Comput. Comput.- Assist. Intervention.*, 2022, pp. 23-33.

[47] S. Lei, T. Gu, Z. Zheng, J. Zhang, "STM-UNet: An efficient u-shaped architecture based on swin transformer and multi-scale MLP for medical image segmentation," *CoRR*, 2023, abs/2304.12615.

[48] H. Wang, P. Cao, Y. Zhang, et al., "Dual-contrastive dual-consistency dual-transformer: A semi-supervised approach to Medical Image segmentation," *in Proc. IEEE/CVF Int. Conf. Comput. Vis*., 2023, pp. 870-879.